\documentclass[aps,showkeys,superscriptaddress,nofootinbib,10pt]{revtex4}
\usepackage[latin1]{inputenc}
\usepackage{graphicx,dcolumn,bm}
\usepackage{subfigure}
\usepackage{color}
\usepackage[colorlinks=true,hyperindex,allcolors=blue]{hyperref}
\usepackage{amsfonts}
\usepackage{amssymb}
\usepackage{amsmath}
\usepackage{latexsym}
\usepackage{times,txfonts}

\usepackage{mathrsfs}

\definecolor{Blue}{rgb}{0.0,0.0,1}
\definecolor{Red}{rgb}{1,0.0,0.0}
\definecolor{Green}{rgb}{0,0.5,0.0}

\begin{document}

\author{Clebson Cruz}
\email{clebson.cruz@ufob.edu.br}
\affiliation{Grupo de Informa\c{c}\~{a}o Qu\^{a}ntica e F\'{i}sica Estat\'{i}stica, Centro de Ci\^{e}ncias Exatas e das Tecnologias, Universidade Federal do Oeste da Bahia - Campus Reitor Edgard Santos. Rua Bertioga, 892, Morada Nobre I, 47810-059 Barreiras, Bahia, Brasil.}

\author{Maron F. Anka}
\email{maronanka@id.uff.br}
\affiliation{Instituto de F\'{i}sica, Universidade Federal Fluminense, Av. Gal. Milton Tavares de Souza s/n, 24210-346 Niter\'{o}i, Rio de Janeiro, Brasil.}

\author{Hamid-Reza Rastegar-Sedehi}
\email{h.rastegar@jahromu.ac.ir}
\affiliation{Department of Physics, College of Sciences, Jahrom University, Jahrom 74135-111, Iran}

\author{Cleidson Castro}\email{ccastro@ufrb.edu.br}
\affiliation {\it Centro de Forma\c c\~ao de Professores, Universidade Federal do Rec\^oncavo da Bahia,Avenida Nestor de Mello Pita, 535 Amargosa, Bahia, Brazil.}

\title{Geometric quantum discord and coherence in a dipolar interacting magnetic system}

\date{\today }

\begin{abstract}
The study of low-dimensional metal complexes has revealed fascinating characteristics regarding the ground-state crossover shown by spin-gaped systems. In this context, this work explores the effect of the quantum-level crossing, induced by the magnetic anisotropies of dipolar interacting systems, on the quantum discord and coherence of the system. The analytical expressions for the quantum discord, based on Schatten 1-norm, and the $l_1$ trace-norm quantum coherence for dinuclear spin-1/2 systems,  are provided in terms of the magnetic anisotropies. The results show that, while the quantum discord has a clear signature of the quantum level-crossing, the basis dependence of the quantum coherence hides the crossover regarding the measured basis. In addition, the global quantum coherence is wholly stored within the correlations of the system, regardless of its reference basis.
\end{abstract}
\keywords{Dipolar Interaction; Qunatum discord; Quantum Coherence; Quantum-level crossing.}

\maketitle

\section{Introduction}

The study of the quantum properties of composite systems has led to a revolution in the development of emerging quantum technologies \cite{mohseni2017commercialize,atzori2019second,PRXQuantum.1.020101}. 
The new generation of quantum devices explores physical properties associated with quantum correlations between particles \cite{cruz2022,CampbellBatteries} and superposition principle for the system states \cite{cruz2020quantifying,caravelli2020random,mohseni2017commercialize}. In this scenario, the characterization of the quantumness of the physical systems is of paramount importance since the existence of quantum correlations and coherence are a valuable resource for several quantum tasks \cite{cruz2022,streltsov2017colloquium,sapienza2019correlations}. 

However, the characterization of quantum correlations is a rather complicated task from the theoretical \cite{huang2014computing} and experimental \cite{cramer} point of view. This scenario is aggravated in Condensed Matter systems, where the number of interacting components in the system is usually on the order of the Avogadro number \cite{mario}. Nevertheless, there are a few exceptions, like low-dimensional metal complexes (LDMC), for which full knowledge about their quantum properties can be obtained through the corresponding analytical solutions \cite{cruz2022,cruz2020quantifying,souza,mario2,cruz2017influence,kova2020unconventional,kuznetsova2013quantum,yuri,yuri2,cruz2022quantum}.  In such solid-state systems, intra-molecular interactions are strong enough to suppress extrinsic and intermolecular interactions \cite{cruz2022,cruz,souza,mario2}.  Therefore, their quantum features exhibit high stability against external perturbations such as temperatures \cite{cruz2022,cruz,souza,mario2,kova2020unconventional}, magnetic fields \cite{cruz2022,cruz2020quantifying,souza2,kova2020unconventional}, and pressures \cite{cruz2017influence,cruz2020quantifying}. These characteristics make these systems promising platforms for the development of emerging quantum technologies \cite{wasielewski2020exploiting,gaita2019molecular,mezenov2019metal,cruz2022,moreno2018molecular}. In regard to these possible applications, the study of dipolar interacting magnetic systems has received considerable attention in the quantum information literature \cite{pinto2018entanglement,kuznetsova2013quantum,pinto2021aspects,castro2016thermal,mohamed2020generation,muthuganesan2021quantum}.

In this work, we present a theoretical study of quantum correlations and coherence for a dipolar interacting magnetic system, exploring the effects of magnetic anisotropies on the quantumness of the system. As a result, this study provides to the literature analytical expressions, in terms of the magnetic anisotropies, for the quantum discord, based on Schatten 1-norm, and the $l_1$ trace-norm quantum coherence written in an arbitrary basis, defined by the co-latitude and longitude angles of the Bloch sphere representation. According to the findings, the behavior of the quantum discord carries a noteworthy signature of the quantum level-crossing, caused by population changes resulting from the alteration of Boltzmann weights arising from the change of the magnetic anisotropies of the system. On the other hand, the basis dependency of quantum coherence is detrimental in terms of recognizing this crossover. In this regard, the measurement of the average quantum coherence is numerically obtained in order to obtain a basis-independent perspective for this quantum resource. The results not only demonstrate that the average coherence is completely stored within the  correlations of the system, yet they also demonstrate that it is possible to retrieve the signature of the energy-level crossover present on the quantum discord measurement.  Furthermore, the findings show how dipolar interaction coupling magnetic anisotropies impact quantum correlations and coherence in a dinuclear spin-1/2 system. Thus, the dipolar interaction model is a viable foundation for quantum technologies based on quantum discord and coherence.

\section{Dinuclear metal complex with Dipolar interaction}
\label{model}

The class of dinuclear metal complexes undergoes several types of magnetic coupling \cite{mario}. Among these are Heisenberg exchange, which is isotropic under rotations in spin space \cite{cruz,cruz2020quantifying,cruz2022,cruz2022quantum}, and Dzyaloshinskii-Moriya (DM) interaction \cite{hoshikawa2021structure,bouammali2021extraction,bouammali2021create}, which accounts for weak ferromagnetism in some antiferromagnetic materials~\cite{mario}. A ubiquitous example of anisotropic coupling in LDMCs is the dipolar interaction \cite{pinto2018entanglement,kuznetsova2013quantum,pinto2021aspects,castro2016thermal,mohamed2020generation,muthuganesan2021quantum}.  This coupling arises from the influence of a magnetic field yielded by one of the magnetic moments in the other  ones~\cite{mario}. In particular, for dinuclear metal complexes, the Hamiltonian which describes this interaction is given by:
\begin{equation}
 \label{eq:hamiltonian001}
\mathcal{H} = -\frac{1}{3}{\vec{S}_A^{T}}\cdot\tensor{D}\cdot{\vec{S}_B} ~,
\end{equation}
where ${\vec{S}_j} = \{ {S_j^x},{S_j^y},{S_j^z}\}$ are the spin operators and $\tensor{D} = \text{diag}(\Delta - 3\epsilon,\Delta + 3\epsilon,-2\Delta)$ is a diagonal tensor, with $\epsilon$ and $\Delta$ being the rhombic and axial parameters, respectively,  related to the magnetic anisotropies in the dipolar model \cite{mario}. In particular,  $\Delta$ is related to the zero-field splitting of the energy levels \cite{moreno2021measuring}. Considering the Hamiltonian, Eq. \eqref{eq:hamiltonian001}, written in the $S^{(z)}$ eigenbasis, $\Delta > 0$ becomes a signature that the spins are on $z$-axis, while $\Delta < 0$ indicates that the spins will be on the $x-y$ plane.

Considering a dinuclear metal complex in with d$^9$ electronic configuration, Eq. \eqref{eq:hamiltonian001} can describe two coupled spin $1/2$ particles in the corresponding $S_z$ eigenbasis $\lbrace \vert 00\rangle,\vert 01\rangle,\vert 10\rangle,\vert 11\rangle\rbrace$
\begin{equation}
\label{Hmatrix}
\mathcal{H} = \frac{1}{6} \left(
\begin{array}{cccc}
 \Delta &  &  & 3\epsilon \\
 & - \Delta & - \Delta &  \\
 & - \Delta & - \Delta &  \\
3\epsilon &  &  &  \Delta
\end{array}\right) ~,
\end{equation}
The energy levels of the system from the coupling parameters are composed by the 
\begin{equation}
\label{eq:energies0001}
\mathcal{E}_{1} = 0 ~,\quad {\mathcal{E}_{2}} = -2\Delta ~,\quad{\mathcal{E}_{{3}}} = \Delta + 3\epsilon ~,\quad{\mathcal{E}_{{4}}} = \Delta - 3\epsilon~.
\end{equation}

From the thermal equilibrium, the density matrix for the coupled system is described by the Gibbs form $\rho_{AB} = {\mathcal{Z}^{-1}}{e^{-\mathcal{H}/{k_B}T}}$, where 
\begin{equation}
\mathcal{Z}= \text{Tr}(e^{-\mathcal{H}/{k_B}T}) =2{e^{{\beta \Delta}/6}}\cosh\left(\frac{\beta \Delta}{6}\right) + 
2{e^{-{\beta \Delta}/6}} \cosh\left(\frac{\beta\epsilon}{2}\right)~.
\end{equation}
is the canonical the partition function, with $k_{B}$ representing the Boltzmann's constant. 
Thus, the dinuclear density matrix at sites labeled by $A$ and $B$ can be written 
in the $S_z$ eigenbasis as  the so-called X-shaped mixed state
\begin{equation}\label{eq:densitymatrix}
\rho_{AB} =\frac{e^{-{\frac{\beta \Delta}{6}}}}{\mathcal{Z}} \left(
\begin{array}{cccc}
\cosh\left(\frac{\beta\epsilon}{2}\right) &  &  & -\sinh\left(\frac{\beta\epsilon}{2}\right) \\
 & {e^{\frac{2\beta \Delta}{6}}}\cosh\left(\frac{\beta \Delta}{6}\right) & {e^{\frac{2\beta \Delta}{6}}}\sinh\left(\frac{\beta \Delta}{6}\right) &  \\
 & {e^{\frac{2\beta \Delta}{6}}}\sinh\left(\frac{\beta \Delta}{6}\right) & {e^{\frac{2\beta \Delta}{6}}}\cosh\left(\frac{\beta \Delta}{6}\right) &  \\
-\sinh\left(\frac{\beta\epsilon}{2}\right) &  &  & \cosh\left(\frac{\beta\epsilon}{2}\right) \\
\end{array}\right) ~.
\end{equation}
The density matrix eigenvalues (population) and their corresponding eigenvectors can be written as:
		\begin{align}
		P_{\Psi^-}&= \frac{1}{1 + e^{\frac{\beta\Delta}{3}} +  2e^{-\frac{\beta\Delta}{6}} \cosh\left(\frac{\beta\epsilon}{2}\right)} \rightarrow |\Psi^{-}\rangle, \label{Ap-x1}\\
		P_{\Psi^+}&=\frac{1}{1 + e^{-\frac{\beta\Delta}{3}} +  2e^{-\frac{\beta\Delta}{6}} \cosh\left(\frac{\beta\epsilon}{2}\right)} \rightarrow |\Psi^{+}\rangle,\label{Ap-x2}\\
		P_{\Phi^+}&=\frac{1}{1 + e^{\beta\epsilon} +  e^{\beta\left(\frac{\Delta+\epsilon}{2}\right)} + e^{\beta\left(\frac{\Delta+3\epsilon}{6}\right)}} \rightarrow |\Phi^{+}\rangle,\label{Ap-x3}\\
		P_{\Phi^-}&=\frac{e^{\beta\epsilon}}{1 + e^{\beta\epsilon} +  e^{\beta\left(\frac{\Delta+\epsilon}{2}\right)} + e^{\beta\left(\frac{\Delta+3\epsilon}{6}\right)}} \rightarrow |\Phi^{-}\rangle, \label{Ap-x4}
		\end{align}
where
\begin{equation}
 \label{bellstates}
|\Psi^{\pm}\rangle = \frac{1}{\sqrt{2}}(|01\rangle \pm |10\rangle) ~,\quad 
|\Phi^{\pm}\rangle = \frac{1}{\sqrt{2}}(|00\rangle \pm |11\rangle) ~.
\end{equation}
are the so-called Bell states, which represent the maximally entangled states for a bipartite system \cite{Nielsen:Book}.

\begin{figure}[h!]
\centering
\includegraphics[scale=0.35]{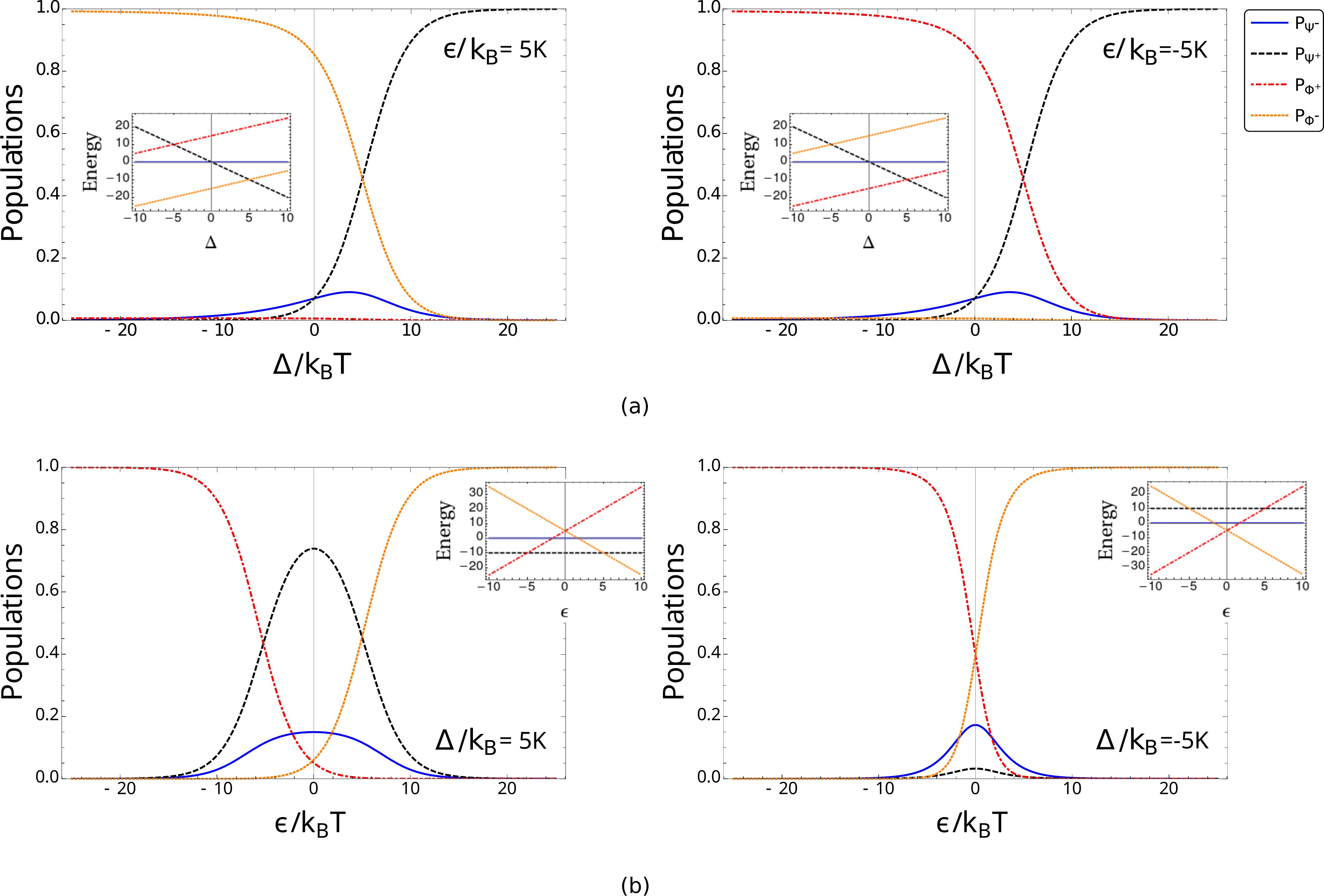}
\caption{(Color online) Populations, Eqs.~\eqref{Ap-x1}--\eqref{Ap-x4}, as a function of the ratio between the magnetic anisotropies and the energy scale factor $k_{B}T$. (a) Axial dependence considering the rhombic parameter  $\epsilon/k_B = 5 $ K (left) and  $\epsilon/k_B = - 5$ K (right). (b) Rhombic dependence considering the axial parameter  $\Delta/k_B = 5 $ K (left) and  $\Delta/k_B = - 5$ K (right). The inset shows the magnetic anisotropy dependence on the energy levels.}
\label{fig:pde}
\end{figure}

The study of LDMC has attracted the attention of both  theoretical and experimental condensed matter physics communities due to the fascinating properties of their ground states ~\cite{chakraborty2019magnetocaloric,cruz2020quantifying}. 
In the presence of an external magnetic field, this systems typically show a quantum level-crossing between its ground state and the first excited one when the field reaches a critical value since the external magnetic field splits its energy levels, changing their corresponding populations. However, since the dipolar interaction arises from the influence of the magnetic field created by one of the magnetic moments in the other, the splitting in energy levels is ruled by the axial ($\Delta$) and rhombic ($\epsilon$) parameters, as can be seen in Eq. \eqref{eq:energies0001}. In this regard, Fig. \ref{fig:pde} shows the  populations as a function of the ratio between the magnetic anisotropies and the energy scale factor $k_{B}T$. 

As can be seen, in agreement with Eq. \eqref{eq:energies0001}, when the spins are in the $x-y$ plane ($\Delta < 0$) with positive rhombic parameter ($\epsilon > 0$), the ground state is the given by the state $|\Phi^{-}\rangle$ (with population $P_{\Phi^-}$), and  there is no quantum level crossing. Thus, the system remains in the  ground state. However, changing the signal of the rhombic parameter induces a quantum level crossing between states $|\Phi^{-}\rangle$ and $|\Phi^{+}\rangle$  (with population $P_{\Phi^+}$). Moreover,  for the spins oriented in the $z$ axis ($\Delta > 0$), it is possible to observe a quantum level crossing between the state $|\Phi^{-}\rangle$ (if $\epsilon > 0$) or $|\Phi^{+}\rangle$ (if $\epsilon < 0$) and the state $|\Psi^{-}\rangle$ (with population $P_{\Psi^+}$) by increasing the ratio $\Delta/k_BT$ to the critical point $\Delta = \vert \epsilon \vert$.

In reference [\onlinecite{castro2016thermal}], the authors study the effect of the magnetic anisotropies, described by the axial and rhombic parameters, on the nonlocal correlations of a dipolar interacting system of two spins-1/2, identified by the Peres-Horodecki separability criterion \cite{horodecki,PhysRevLett.77.1413}. In addition, they explore the change in the ground state on the thermal entanglement for the teleportation process. However, although quantum entanglement provides one path toward the characterization of nonlocal  correlations, it does not encompass all quantum correlations in the system \cite{zurek,vedral4,liu,ma,adesso2,adesso,cruz,sarandy,paula,sarandy3,luo,datta,vedral}. Therefore, in order to expand this result, the following section presents a study of the quantum correlations and coherence described by the Schatten 1-norm geometric quantum discord and the $l_1$ trace-norm quantum coherence.  

\section{Quantum Discord}

Quantum discord has been defined as a measurement of the quantumness of correlations in a quantum system. It has been first introduced as an entropic measurement of genuinely quantum correlations in a quantum state, defined as the difference between the total and the classical correlation \cite{zurek} $\mathcal{Q}(\rho_{AB})=\mathcal{I}(\rho_A:\rho_B)-\mathcal{C}(\rho_{AB})$,
where $\mathcal{I}(\rho_A:\rho_B)=S(\rho_A)+S(\rho_B)-S(\rho_{AB})$ represents the mutual information between the subsystems $A$ and $B$, and $\mathcal{C}(\rho_{AB})$ is the classical correlation of the composite system $\rho_{AB}$ defined as $\mathcal{C}(\rho_{AB})=\max_{\lbrace B_k\rbrace} \left[ S(\rho_A)-\sum_{k} p_kS\left(\rho_k\right)\right]$, with the maximization taking over positive operator-valued measurements (POVM's) $\lbrace B_k\rbrace$ performed locally only on subsystem $B$.
However, this analytical maximization over POVMs is an arduous task even for a two-qubit system \cite{paula,zurek,vedral,vedral4,huang2014computing,cramer}. In this scenario, the class of entropic measurements of correlations, such as the entropic quantum discord, is defined as nondeterministic polynomial time (NP-complete) problems \cite{huang2014computing}. Consequently, only a few results for the analytical expression of entropic quantum discord, and only for certain classes of states are exact solutions known \cite{ma,luo,sarandy,adesso,datta,vedral4,terno}.  Due to this fact, alternative measurements of quantum correlations have been proposed \cite{ma,luo,obando,sarandy,adesso,datta,vedral4,terno,girolami,girolami2,girolami2014quantum,vedral5,paula,sarandy3,piani,paula2,spehner2016geometric}, especially quantifiers based on geometric arguments \cite{luo,obando,vedral5,paula,sarandy3,piani,paula2,spehner2016geometric}.

Geometric approaches are widely used to characterize and quantify quantum resources in a wide variety of quantum systems \cite{hu2018quantum}. In particular, the  Schatten 1-norm quantum discord \cite{paula,sarandy3,cruz,cruz2022,khedif2022non}, is a reliable geometric-based quantifier of the amount of quantum correlations in metal complexes \cite{cruz,cruz2016quantum,cruz2022,khedif2022non}. The so-called geometric quantum discord can be defined in terms of the minimal distance between a set $\omega$ of closest classical-quantum states $\rho_c$ \cite{paula,sarandy3,cruz}, given by:
\begin{equation}
\rho_c=\sum_{k}p_{k}\Pi_{k}^{\lbrace A \rbrace}\otimes\rho_{k}^{\lbrace B \rbrace},
\label{eq:01}
\end{equation}
where $0 \leq p_k \leq 1$ and $\sum_{k} p_k = 1$; $\lbrace\Pi_{k}^{\lbrace A \rbrace}\rbrace$ define a set of orthogonal projectors for a given subsystem $A$ and $\rho_{k}^{\lbrace B \rbrace}$ the reduced density matrix for the subsystem $B$ \cite{paula,sarandy3}.
Therefore, the geometric quantum discord can be expressed as 
\begin{eqnarray}
\mathcal{Q}_G(\rho_{AB})=\min_{\omega}\Vert\rho_{AB} - \rho_c\Vert~,\label{eq:02}
\end{eqnarray}
where $\Vert M\Vert=\mbox{Tr}\left[\sqrt{M^\dagger M}\right]$ is the so-called 1-norm, and $\rho_{AB}$ is the given quantum state at thermal equilibrium, Eq. \eqref{eq:densitymatrix}.

Therefore, considering the given dinuclear magnetic system of spins-$1/2$ in a quantum spin-lattice, ruled by a dipolar Hamiltonian $\mathcal{H}$, Eq.\eqref{eq:hamiltonian001}, the invariance under $\pi$ rotation around a given spin axis ($\mathbb{Z}_2$ symmetry) \cite{mario,sarandy} 
allow us to compute the geometric quantum discord, based on Schatten 1-norm, for the two-qubit X state, Eq. \eqref{eq:densitymatrix}, as \cite{ciccarello,obando}
\begin{eqnarray}
\mathcal{Q}_G(\rho_{AB}) = \frac{1}{2}\sqrt{\frac{\phi_1^2\mbox{max}\lbrace{\phi_2^2,\phi_3^2}\rbrace-\phi_2^2\mbox{min}\lbrace{\phi_{1}^2,\phi_{3}^2}\rbrace}{\mbox{max}\lbrace{\phi_2^2,\phi_3^2}\rbrace-\mbox{min}\lbrace{\phi_{1}^2,\phi_{3}^2}\rbrace+\phi_1^2-\phi_2^2}}
\label{discord}
\end{eqnarray}
where
\begin{eqnarray}
\phi_{1} &=& \frac{e^{\frac{\beta\Delta}{6}}\left|
   -1+e^{\beta\Delta/3}\right| +2\left|\sinh \left(\frac{\beta\epsilon}{2}\right)\right| }{\left| 2 \cosh   \left(\frac{\beta\epsilon}{2}\right)+e^{\beta\Delta/6} 
   +e^{\beta\Delta/2}\right|},  \\
\phi_{2} &=& \frac{e^{\frac{\Delta}{6}}\left| -1+e^{\beta\Delta/3}\right|-2
   \left| \sinh \left(\frac{\beta\epsilon}{2}\right)\right| }{\left| 2 \cosh
   \left(\frac{\beta\epsilon}{2}\right)+e^{\beta\Delta/6}+e^{\beta\Delta/2}\right| }, \\ 
\phi_{3} &=& \frac{2}{e^{\beta\Delta/3} \cosh \left(\frac{\beta\Delta}{6}\right)
   \text{sech}\left(\frac{\beta\epsilon}{2}\right)+1}-1~.
\end{eqnarray}

\begin{figure}[ht]
\centering
\includegraphics[scale=0.2]{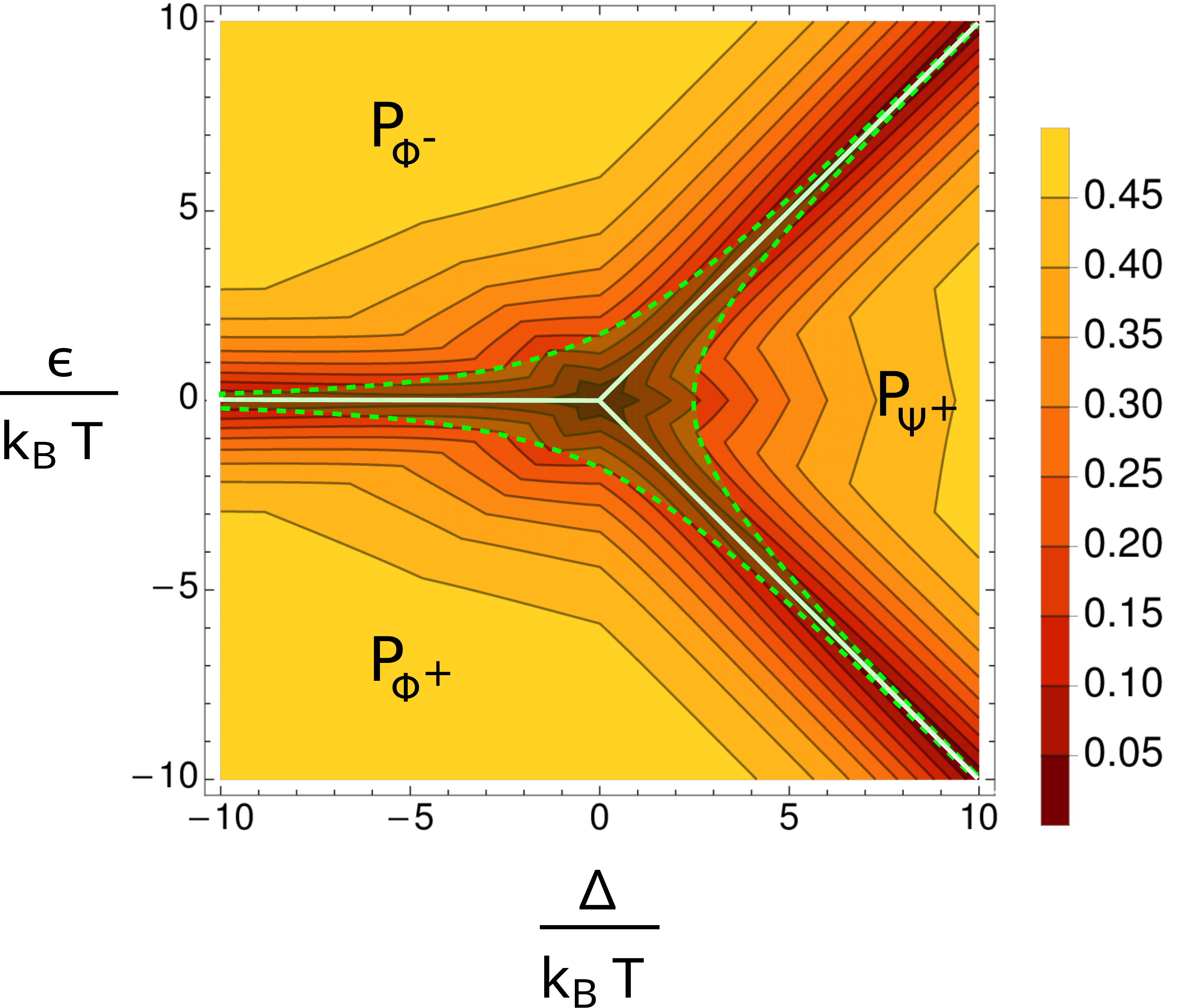}
\caption{(Color online) Quantum Discord, based on Schatten 1-norm, for a dipolar interacting magnetic system, Eq. \eqref{discord}, as a function of the ratios $\Delta/{k_B}T$ and $\epsilon/{k_B}T$. The solid white line denotes the boundary between the quantum level crossings. The dashed green line is the boundary given by the concurrence, Eq. \eqref{concurrence}, $\mathbb{C} (\rho_{AB} )=0$, inside which the entanglement of the system is absent.}
\label{fig2}
\end{figure}

Considering the dipolar magnetic system in thermal equilibrium described by Eq. \eqref{eq:densitymatrix}, it is possible to examine how the magnetic anisotropies, represented by the axial ($\Delta$) and rhombic ($\epsilon$) coupling parameters, affects the thermal quantum discord in the system. Fig. \ref{fig2} shows the geometric quantum discord, based on Schatten 1-norm, Eq. \eqref{discord}, as a function of the ratio $\Delta/{k_B}T$ and $\epsilon/{k_B}T$. As expected, the quantum discord reaches its maximum (saturated) value of 1/2 as T approaches zero. As the temperature rises, the value of quantum discord decreases inexorably and goes to zero when $T \gg \vert\Delta\vert$ and $T\gg \vert\epsilon\vert$. On the other hand, given the spins in the $x-y$ plane ($\Delta < 0$), it is sufficient that only $T\gg \vert\epsilon\vert$ to the discord reaches its minimum value. However, if the spins are in the  $z$-axis  ($\Delta > 0$), one can increase the quantum discord by increasing the axial parameter $\Delta$ even when  $T\gg \vert\epsilon\vert$.

Furthermore, regarding the magnetic anisotropies, the quantum discord presents a signature of the quantum level crossing in the dipolar interacting system, highlighted on the solid white line in Fig. \ref{fig2}. 
Considering the spins oriented in the $z$-axis ($\Delta > 0$), the zero-field splitting leads the system to a quantum level crossing in the critical boundary $\Delta = \vert \epsilon \vert$, where it is possible to detect a crossover between the states $|\Psi_{+}\rangle$ and $|\Phi_{-}\rangle$, if $\epsilon > 0$, or  $|\Phi_{+}\rangle$, if $\epsilon < 0$. Moreover,  for the spins oriented in the $x-y$ plane ($\Delta < 0$), it is possible to observe a quantum level crossing between the state $|\Phi^{-}\rangle$ and $|\Phi^{-}\rangle$ in the critical boundary $\epsilon = 0$. On the other hand, the degree of quantum discord in the system can be increased by gradient ascent of the function $\mathcal{Q}_G(\rho_{AB})$, perpendicularly to the crossing boundary, which occurs for values in which $\vert \epsilon \vert \gg {k_B}T$ (for $\Delta < 0$), $\Delta\gg {k_B}T$ (for $\vert \epsilon \vert \ll \Delta$), and $\vert \epsilon \vert \gg \Delta$), corresponding to the lightest region in Fig. \ref{fig2}.
Therefore, by controlling the axial ($\Delta$) and rhombic ($\epsilon$) anisotropies is possible to manage the degree of quantum discord in the dipolar interacting system.


In addition, in order to compare quantum discord to the level of entanglement in the system under investigation, we use the concurrence measure. Typically, concurrence is used to assess entanglement in bipartite systems, and it can be easily computed for any two-qubit system. The thermal concurrence examines the resemblance between the considered quantum state in thermal equilibrium and its bit-flipped density matrix, $\Bar{\rho} = \rho_{AB}  (\sigma^y \otimes \sigma^y) \rho_{AB}  ^{*} (\sigma^y \otimes \sigma^y)$. In particular, for the X- shaped density matrix, Eq. \eqref{eq:densitymatrix}, the concurrence is analytically defined as
\begin{equation}
\mathbb{C} (\rho_{AB} ) := \rm{max} \{0, \mathcal{A}, \mathcal{B}\},
\label{concurrence}
\end{equation}
where
\begin{eqnarray}
    \mathcal{A} &=& \frac{e^{-{\frac{\beta \Delta}{6}}}}{\mathcal{Z}}\left[\left| {e^{\frac{2\beta \Delta}{6}}}\sinh\left(\frac{\beta \Delta}{6}\right)\right| - \cosh\left(\frac{\beta\epsilon}{2}\right) \right]~,\\
    \mathcal{B} &=& \frac{e^{-{\frac{\beta \Delta}{6}}}}{\mathcal{Z}}\left[\left|\sinh\left(\frac{\beta\epsilon}{2}\right)\right| -  {e^{\frac{2\beta \Delta}{6}}}\cosh\left(\frac{\beta \Delta}{6}\right)\right]~.
\end{eqnarray}

Dashed green line in Fig. \ref{fig2} denotes the boundary given by $\mathbb{C} (\rho_{AB} )=0$. Inside this region, the concurrence is zero, and the state of the system is separable. However, within the region where entanglement is absent, the quantum discord of the system is still considerably more than zero, ensuring the presence of quantum-correlated states even when the system is in a separable syaye. On the other hand, for low temperatures, the entanglement is zero in the quantum level crossing boundary alongside the quantum discord at the quantum level crossing boundary. In this scenario, the existence or absence of entanglement and, therefore, quantum correlations, is dependent on its ground state, which might vary in response to magnetic anisotropies. Thus, the variation of Boltzmann's weights, Eqs. \eqref{Ap-x1}-\eqref{Ap-x4}, associated with the occupancy of the energy levels, is the physical mechanism responsible for the abrupt change in the quantum correlations near the energy-level crossover.

\section{Quantum Coherence}

Similar to the approach proposed for the entanglement theory, where the quantum entanglement can be characterized by the distance between a state of interest ($\rho$) and a set of states closed under local operations, and classical communication (separable states) \cite{baumgratz2014quantifying,hu2018quantum,horodecki}, Baumgratz et al. \cite{baumgratz2014quantifying} provided the mathematical tools for quantifying the amount  of quantum coherence in a quantum system. Considering a $d$-dimensional Hilbert space, quantum coherence can be obtained from the minimal value of a distance measurement $D(\rho,\sigma)$, between the considered quantum state $\rho$ and a set $\lbrace \sigma=\sum_{k}^{d} \vert k\rangle\langle k \vert \in \mathcal{I} \rbrace$ of incoherent states, 
where the reference basis $\lbrace \vert k\rangle \rbrace_{\{k=1,...,d\}}$ can be adequately defined considering the physics of the problem under investigation or the task that requires  this quantum resource \cite{streltsov2016quantum,baumgratz2014quantifying,cruz2020quantifying}. 
In this scenario, since the non-vanishing off-diagonal terms of the density operator $\rho$, which characterizes the quantum state of the system of interest, constitute the superposition from the chosen reference basis \cite{hu2018quantum,baumgratz2014quantifying}, the authors   established a reliable measurement of quantum coherence through the $l_{1}$ trace norm as \cite{baumgratz2014quantifying}
\begin{eqnarray}
\mathcal{C}_{l_{1}}&=& \min_{\sigma \in \mathcal{I}} \Vert \rho -\sigma \Vert_{l_1}=\sum_{i\neq j} \vert \langle i\vert \rho\vert j \rangle\vert~.
\label{coherence}
\end{eqnarray}

Since coherence is a quantity that is reliant on the basis on which it is measured, it is essential to choose a reference basis for the system within a metrology setting  \cite{hu2018quantum,baumgratz2014quantifying}. In this scenario, the basis of an arbitrary quantum state can be altered by means of unitary operations \cite{hu2018quantum,Nielsen:Book}. In particular, for two-level systems such as spin-1/2, any reference basis can be obtained from the unitary transformation
\begin{equation}
    \mathcal{U}(\theta,\phi) = \left(
\begin{array}{cc}
 \cos \left(\frac{\theta }{2}\right) & -e^{i \phi } \sin \left(\frac{\theta }{2}\right) \\
 e^{-i \phi } \sin \left(\frac{\theta }{2}\right) & \cos \left(\frac{\theta }{2}\right) \\
\end{array}
\right)~,
\end{equation}
where the $\theta$ and $\phi$ angles are the spherical equivalents of the co-latitude with respect to the z-axis, and the longitude concerning the x-axis in a Bloch sphere representation, respectively \cite{Nielsen:Book,rojas}. In this regard, the unitary transformation for the bipartite state given by Eq. \eqref{eq:densitymatrix} is given by $\rho_{AB}^{\lbrace \theta,\phi\rbrace} = \hat{\mathcal{U}}_{AB}(\theta,\phi)\rho_{AB}\hat{\mathcal{U}}_{AB}(\theta,\phi)$ \cite{rojas}, where
\begin{equation}
    \hat{\mathcal{U}}_{AB}(\theta,\phi) = \mathcal{U}(\theta,\phi)\otimes \mathcal{U}(\theta,\phi) =\left(
\begin{array}{cccc}
 \cos ^2\left(\frac{\theta }{2}\right) & -e^{i \phi } \sin \left(\frac{\theta }{2}\right) \cos \left(\frac{\theta }{2}\right) & -e^{i \phi } \sin
   \left(\frac{\theta }{2}\right) \cos \left(\frac{\theta }{2}\right) & e^{2 i \phi } \sin ^2\left(\frac{\theta }{2}\right) \\
 e^{-i \phi } \sin \left(\frac{\theta }{2}\right) \cos \left(\frac{\theta }{2}\right) & \cos ^2\left(\frac{\theta }{2}\right) & -\sin
   ^2\left(\frac{\theta }{2}\right) & -e^{i \phi } \sin \left(\frac{\theta }{2}\right) \cos \left(\frac{\theta }{2}\right) \\
 e^{-i \phi } \sin \left(\frac{\theta }{2}\right) \cos \left(\frac{\theta }{2}\right) & -\sin ^2\left(\frac{\theta }{2}\right) & \cos
   ^2\left(\frac{\theta }{2}\right) & -e^{i \phi } \sin \left(\frac{\theta }{2}\right) \cos \left(\frac{\theta }{2}\right) \\
 e^{-2 i \phi } \sin ^2\left(\frac{\theta }{2}\right) & e^{-i \phi } \sin \left(\frac{\theta }{2}\right) \cos \left(\frac{\theta }{2}\right) &
   e^{-i \phi } \sin \left(\frac{\theta }{2}\right) \cos \left(\frac{\theta }{2}\right) & \cos ^2\left(\frac{\theta }{2}\right) \\
\end{array}
\right)
\label{unitary}
\end{equation}    

By varying the co-latitude and longitude angles ${\lbrace \theta,\phi\rbrace}$, one can obtain the bipartite state $\rho_{AB}$, Eq. \eqref{eq:densitymatrix}, in any reference basis. Using the unitary transformation for the bipartite states, Eq. \eqref{unitary}, in Eq. \eqref{eq:densitymatrix}, one can obtain the representation of the density operator for the dipolar interacting magnetic system of two spins-1/2 written in an arbitrary basis as
\begin{equation}
\rho_{AB}^{\lbrace \theta,\phi\rbrace} = \frac{e^{-{\frac{\beta \Delta}{6}}}}{4\mathcal{Z}} \left(
\begin{array}{cccc}
 \varrho_{11} & \varrho_{12} & \varrho_{12} & \varrho_{14} \\
\varrho_{12}^* & \varrho_{22} & \varrho_{23} & -\varrho_{12} \\
\varrho_{12}^* & \varrho_{23} & \varrho_{22} & -\varrho_{12} \\
\varrho_{14}^* & -\varrho_{12}^* & -\varrho_{12}^* & \varrho_{11} \\
\end{array}
\right)~,
\label{eq:densitymatrixarb}
\end{equation}
where
\begin{eqnarray}
    \varrho_{11} &=& 2 \sin ^2(\theta ) \left(\sinh \left(\frac{\beta \Delta }{2}\right)+\cosh \left(\frac{\beta \Delta }{2}\right)-\sinh \left(\frac{\beta \epsilon }{2}\right) \cos (2 \phi )\right)+\cosh \left(\frac{\beta \epsilon}{2}\right) (\cos (2 \theta )+3)~,\\
    \varrho_{12} &=& e^{-3 i \phi } \sin (\theta ) \left(2 e^{2 i \phi } \cos (\theta ) \left(\cosh \left(\frac{\beta \epsilon }{2}\right)-e^{\beta \Delta /2}\right)+e^{4 i \phi } \sinh \left(\frac{\beta \epsilon }{2}\right) (\cos (\theta
   )+1)+\sinh \left(\frac{\beta \epsilon }{2}\right) (\cos (\theta )-1)\right)~,\\
   \varrho_{14} &=& 2 e^{-2 i \phi } \sin ^2(\theta ) \left(\cosh \left(\frac{\beta \epsilon }{2}\right)-e^{\beta \Delta /2}\right)-4 e^{-4 i \phi } \sinh \left(\frac{\beta \epsilon }{2}\right) \sin ^4\left(\frac{\theta }{2}\right)-4 \sinh
   \left(\frac{\beta \epsilon }{2}\right) \cos ^4\left(\frac{\theta }{2}\right)~,\\
    \varrho_{22} &=& 2 \left(e^{\beta \Delta /2} \cos ^2(\theta )+e^{\beta \Delta /6}+\sin ^2(\theta ) \left(\sinh \left(\frac{\beta \epsilon }{2}\right) \cos (2 \phi )+\cosh \left(\frac{\beta \epsilon }{2}\right)\right)\right)~,\\
    \varrho_{23} &=& 2 e^{\beta \Delta /2} \cos ^2(\theta )-2 e^{\beta \Delta /6}+2 \sin ^2(\theta ) \left(\sinh \left(\frac{\beta \epsilon }{2}\right) \cos (2 \phi )+\cosh \left(\frac{\beta \epsilon }{2}\right)\right)~.
\end{eqnarray}

The diagonal entries of Eq. \eqref{eq:densitymatrixarb} are real, and the trace is $1$. In addition, to ensure real eigenvalues, hermiticity restricts off-diagonal elements to two complex numbers, i.e., $\varrho_{ij}$ is the complex conjugate of $\varrho_{ji}$. 

Thus, from Eqs.  \eqref{coherence} and \eqref{eq:densitymatrixarb}, it is possible to write an analytical expression for the normalized quantum coherence in an arbitrary basis, defined by the co-latitude and longitude angles ${\lbrace \theta,\phi\rbrace}$, as:
\begin{equation}
    \mathcal{C}_{l_1} ^{\lbrace \theta,\phi\rbrace} =  \frac{e^{-{\frac{\beta \Delta}{6}}}}{6\left|\mathcal{Z}\right|} \left[ 4\left| \varrho_{12} \right| +  \left| \varrho_{14} \right| +  \left| \varrho_{23} \right| \right]~.
    \label{coherence_arb}
\end{equation}

In order to examine the relationship between quantum coherence and quantum correlations, a new metric known as correlated coherence was established recently \cite{kraft2018genuine,rojas,tan2016unified}. Quantum correlated coherence is a measure of coherence in which all local components have been eliminated, i.e., all coherence in the system is totally recorded in the quantum correlations. For any given quantum state $\rho$, the correlated contribution to quantum coherence may be calculated by subtracting the local coherence of subsystems $\rho_A = \text{Tr}_B (\rho)$ and $\rho_B = \text{Tr}_A (\rho)$ from the overall coherence \cite{kraft2018genuine,tan2016unified}. Thus, the definition of correlated coherence according to the $l_1$-norm of coherence is:
\begin{equation}
\mathcal{C}_{cc} (\rho_{AB}^{\lbrace \theta,\phi\rbrace} ) := \mathcal{C}_{l_1} (\rho)-\mathcal{C}_{l_1} (\rho_A)-\mathcal{C}_{l_1} (\rho_B).
\label{correlated_coherence}
\end{equation}

Considering the density matrix of the dipolar interacting magnetic system written in an arbitrary basis, Eq. \eqref{eq:densitymatrixarb}, the reduced density matrices of local subsystems are $\rho_A = \rho _B = \mathbb{I}/2$, the maximally mixed state. Thus, regardless of the basis, the local subsystems will remain in the maximally mixed state, since it is basis invariant \cite{Nielsen:Book}. Consequently, the local contribution for the quantum coherence in this dipolar interacting system is always null, and the global coherence of the system, Eq. \eqref{coherence_arb}, is totally recorded in the quantum correlations of the system, regardless of its reference basis. Therefore, for a number of different combinations of values for the co-latitude and longitude angles ${\lbrace \theta,\phi\rbrace}$, the unitary transformation, Eq. \eqref{unitary}, gives a direct connection between the overall and the correlated degrees of coherence. 

\subsection{Axial Coherence}

In particular, due to the rotation symmetry of the dipolar interaction, the density matrix will be invariant when rotated both spins by an angle $\pi$ along any given spin axis. Thus, choosing the co-latitude angle as $\theta=n\pi$ ($n=\lbrace 0, 1, 2,...\rbrace$), regardless of the longitude angle $\phi$, one can obtain the density matrix in the X-shaped form as described in Eq. \eqref{eq:densitymatrix}. On the other hand, by applying the unitary transformation for the bipartite states, Eq. \eqref{unitary}, for $\lbrace\theta=\pi/2;\phi = n\pi\rbrace$,  and $\lbrace\theta=\pi/2;\phi = n\pi/2\rbrace$, in Eq. \eqref{eq:densitymatrix} one can obtain the density matrix $S^{(x)}$ and $S^{(y)}$ eigenbasis, respectively. 
\begin{equation}
\rho_{AB}^{\lbrace X,Y\rbrace} = \frac{e^{-{\frac{\beta \Delta}{6}}}}{2\mathcal{Z}} \left(
\begin{array}{cccc}
 e^{\beta\Delta/2}+e^{\mp\beta\epsilon/2} & 0 & 0 & \mp\left(e^{\beta\Delta/2}-e^{\mp \beta\epsilon/2}\right) \\
 0 & e^{\beta\Delta/6}+e^{\pm \beta\epsilon/2} & e^{\pm \beta\epsilon/2}-e^{\beta\Delta/6} & 0 \\
 0 & e^{\pm \beta\epsilon/2}-e^{\beta\Delta/6} & e^{\beta\Delta/6}+e^{\pm \beta\epsilon/2} & 0 \\
 \mp\left(e^{\beta\Delta/2}-e^{\mp \beta\epsilon/2}\right) & 0 & 0 & e^{\beta\Delta/2}+e^{\mp{\beta\epsilon/2}} \\
\end{array}
\right)~.
\label{eq:densitymatrixxy}
\end{equation}

As can be seen, due to the symmetry of the X-shaped density matrices \cite{rau2009algebraic}, the X-structure of the operator is preserved. Therefore, from Eqs. \eqref{eq:densitymatrix}, \eqref{coherence} and \eqref{eq:densitymatrixxy}, one can obtain the analytical expressions for the normalized axial quantum coherences as
\begin{eqnarray}
    \mathcal{C}_{l_1} ^{\lbrace Z\rbrace} &=&  \frac{2}{3\left| \mathcal{Z}\right|}\left[ e^{\beta\Delta/6} \left|\sinh \left(\frac{\beta\Delta}{6}\right)\right| + e^{-\beta\Delta/6} \left| \sinh \left(\frac{\beta\epsilon}{2}\right)\right|\right]
    \label{zcoherence}\\
\mathcal{C}_{l_1}^{\lbrace X,Y\rbrace} &=& \frac{e^{-{\frac{\beta \Delta}{6}}}}{3\left|\mathcal{Z}\right|} \left[\left| e^{\beta\Delta/2}-e^{\mp\beta\epsilon/2}\right| +\left| e^{\beta\Delta/6}-e^{\pm\beta\epsilon/2}\right| \right]    \label{xycoherence}
\end{eqnarray}

\begin{figure}[ht]
    \centering
\includegraphics[scale=0.4]{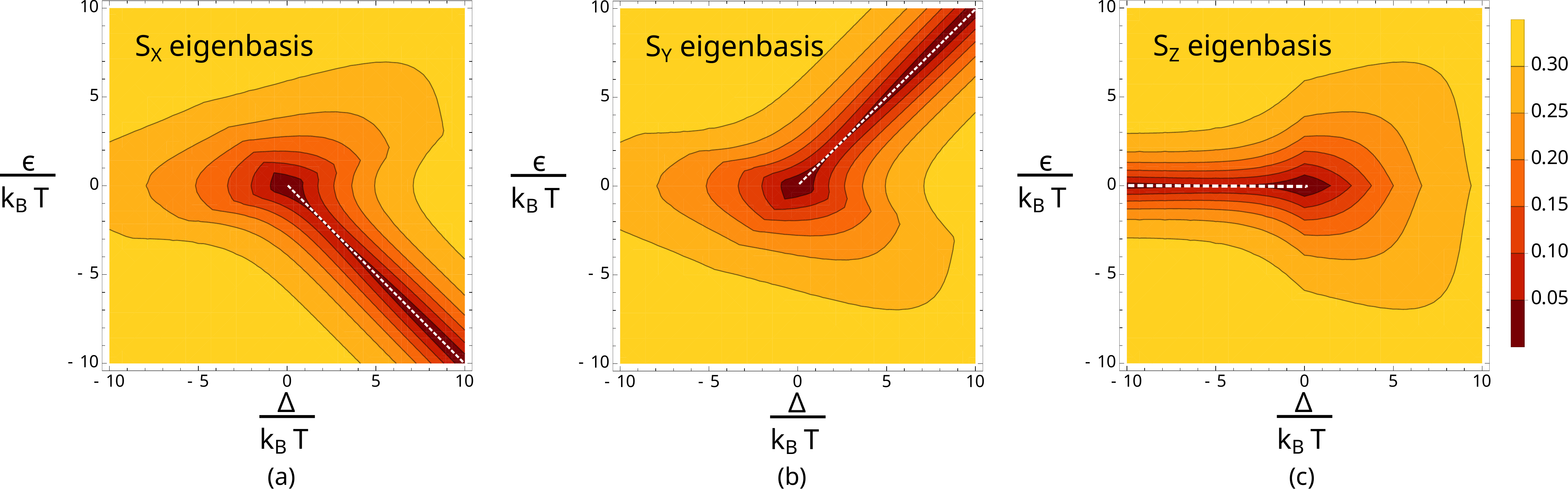}
\caption{(Color online) Axial quantum coherence based on $l_{1}$ trace norm, for a dipolar interacting magnetic system, as a function of the ratios $\Delta/{k_B}T$ and $\epsilon/{k_B}T$. The dashed white line represents the minimum value for the axial quantum coherence.}
\label{fig3}
\end{figure}

Fig. \ref{fig3} shows the axial quantum coherence in $S^{(i)}$ spin eigenbasis, where $i=\lbrace x,y,z\rbrace$. Different from the behavior observed for quantum discord (see Fig. \ref{fig2}), the axial quantum coherence is not sensible to the quantum level crossing.  The quantum coherence in each axis $\lbrace x,y,z\rbrace$ is minimized in only one energy-level crossover. As can be seen, considering the spins oriented in the $z$-axis  ($\Delta > 0$), the axial coherence in the $S^{(x)}$ eigenbasis is minimized on the critical boundary $\Delta = -\epsilon$ (with $\epsilon < 0$), where it is possible to detect a crossover between the states $|\Psi_{+}\rangle$ and $|\Phi_{+}\rangle$, while the coherence in $S^{(y)}$ eigenbasis is minimized on the critical boundary $\Delta = \epsilon$ (with $\epsilon > 0$), where it is possible to detect a crossover between the states $|\Psi_{+}\rangle$ and $|\Phi_{-}\rangle$  (see Fig. \ref{fig2}). As expected from Eq. \ref{xycoherence}, if the rhombic parameter is null ($\epsilon = 0$), $\mathcal{C}_{l_1}^{\lbrace X\rbrace}=\mathcal{C}_{l_1}^{\lbrace Y\rbrace}$. 
On the other hand,  for the spins oriented in the $x-y$ plane, ($\Delta < 0$), it is possible to observe that $\mathcal{C}_{l_1} ^{\lbrace Z\rbrace}$ is minimized in the quantum level crossing between the state $|\Phi^{-}\rangle$ and $|\Phi^{-}\rangle$ in the critical boundary $\epsilon = 0$.  

As shown in Fig. \ref{fig3}, the basis dependence of the quantum coherence hides the energy-level crossover in this dipolar interacting system regarding the measured basis. Therefore, the basis dependence on the quantum coherence defined by Baumgratz et al. \cite{baumgratz2014quantifying}, can be unfavorable to recognizing the quantum level crossing caused by population changes resulting from the alteration of Boltzman weights, Eqs. \eqref{Ap-x1}-\eqref{Ap-x4}, arising from the change of the magnetic anisotropies of the dipolar interacting system.

\subsection{Average Coherence}

Since the coherence formulated in the quantum resource theory is a basis-dependent measurement  \cite{hu2018quantum,streltsov2016quantum,streltsov2017colloquium}, it is natural to define a basis-independent measurement \cite{liu2023average,luo2019average,cheng2015complementarity,yao2015quantum,designolle2021set}. Recent research has shown, via the use of relative entropies, as distance measurements of quantum correlations, that basis-independent measurements of entropic quantum coherence are precisely identical to entropic discord \cite{yao2015quantum}. On the other hand, a possible basis-free measurement of quantum coherence for a quantum system can be obtained from a geometrical standpoint by averaging the coherence of a state across all reference bases \cite{liu2023average,luo2019average,cheng2015complementarity,designolle2021set}. From a theoretical point of view, this measurement corresponds to averaging the coherence on a standard basis across all equivalent states $\rho_{AB}^{\lbrace \theta,\phi\rbrace} = \hat{\mathcal{U}}_{AB}(\theta,\phi)\rho_{AB}\hat{\mathcal{U}}_{AB}(\theta,\phi)$. Therefore, as any two-qubit reference base can be created by applying the unitary operation described in Eq. \eqref{unitary}, the average quantum coherence can be obtained from Eq. \eqref{coherence_arb} as
\begin{equation}
    \langle \mathcal{C}_{l_1} \rangle = \frac{1}{4\pi}\int\limits_{0}^{2\pi} \int\limits_{0}^{\pi} \sin{\left(\theta\right)} \mathcal{C}_{l_1} ^{\lbrace \theta,\phi\rbrace} d\theta d\phi~.
    \label{average}
\end{equation}

It is worth mentioning that these integrals are not trivial to solve, and an analytical expression for the average coherence is not presented. However, it can be numerically integrated by any quadrature method \cite{press2007numerical}. In this scenario, Eq. \eqref{average}  is estimated by using the Clenshaw-Curtis rule on adaptively refined subintervals of the integration area \cite{liu2019clenshaw,press2007numerical} since the numerical integration algorithms are often equally efficient and effective as conventional algorithms for well-behaved integrands such as Eqs. \eqref{coherence_arb} and \eqref{correlated_coherence} \cite{press2007numerical}. 

\begin{figure}[ht]
\centering
\includegraphics[scale=0.5]{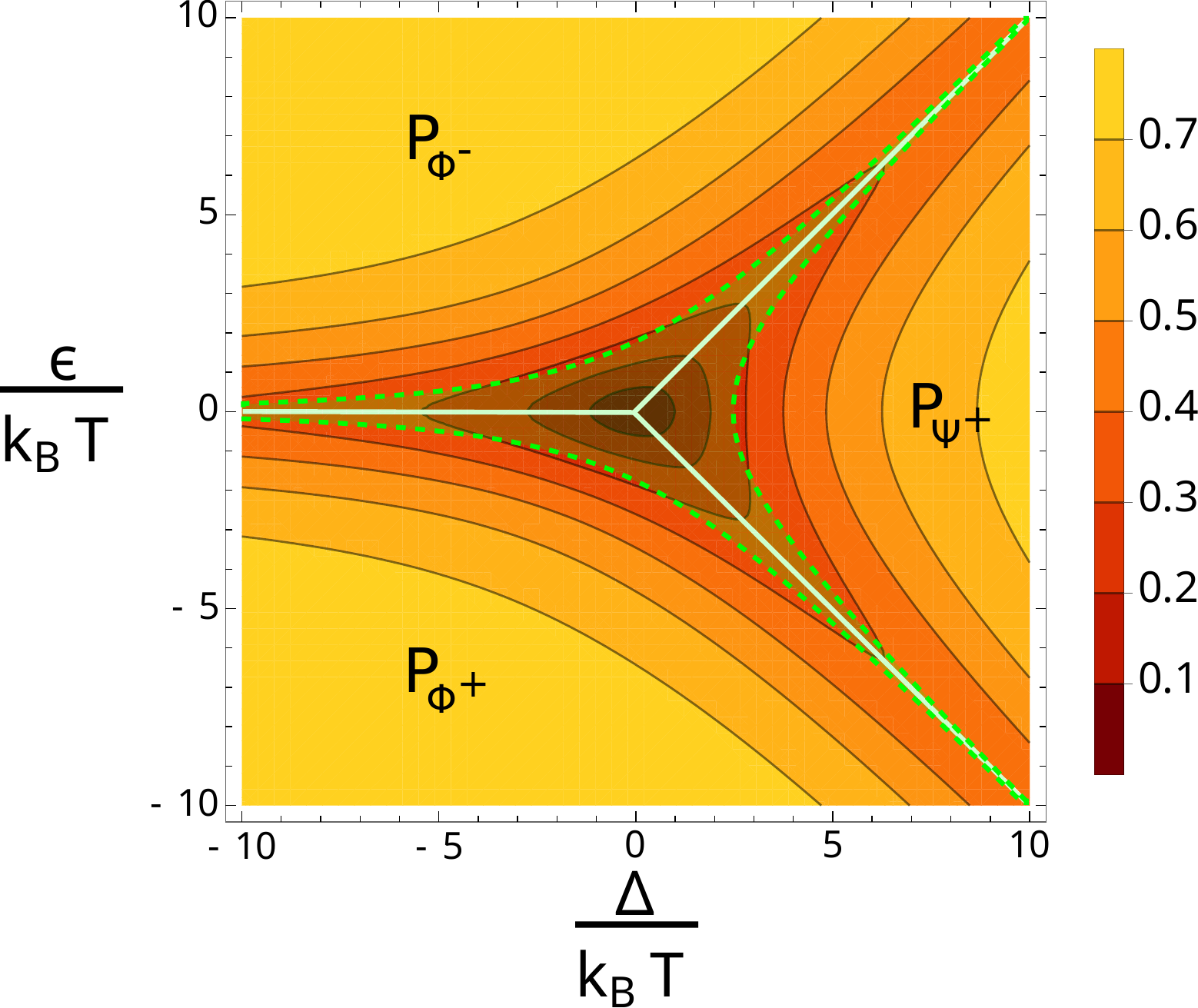}
\caption{(Color online) Average quantum coherence based on  $l_{1}$ trace norm, for a dipolar interacting magnetic system, as a function of the ratios $\Delta/{k_B}T$ and $\epsilon/{k_B}T$.  The solid white line represents the boundary between the quantum level crossings. The dashed green line is the boundary given by the concurrence, Eq. \eqref{concurrence}, $\mathbb{C} (\rho_{AB} )=0$, inside which the entanglement of the system is absent.}
\label{fig4}
\end{figure}

Fig. \ref{fig4} shows the average quantum coherence for the dipolar magnetic interacting system. The solid white line represents the threshold at which the quantum-level crossing, described in previous sections, actually occurs. As expected, based on  Fig. \ref{fig3}, when the temperature rises reaching the threshold $T \gg \vert\Delta\vert$ and $T\gg \vert\epsilon\vert$, the value of coherence reaches its lowest point and will be equal to zero. However, the behavior of the average coherence is completely different from that observed in the axial (basis-dependent) coherence shown in Fig. \ref{fig3}. 

Moreover, besides unified frameworks from relative entropic measurements has shown that basis-independent entropic quantum coherence is equivalent  to entropic discord \cite{yao2015quantum}, this is not true for this geometrical approach. However, although the contour lines of the average coherence are quite different from that shown in the discord presented in Fig. \ref{fig2}, it is still able to identify the signature of the energy-level crossing that was seen during the measurement of the quantum discord. This result is due to the fact that the global coherence is totally stored within the correlations of the system, and its average behavior is affected by the presence of genuine quantum correlations measured by the quantum discord.

In addition, the entanglement of the system is absent within the area shown by the dashed green line that denotes the boundary supplied by the concurrence, which is denoted by Eq. \eqref{concurrence}, $\mathbb{C} (\rho_{AB})=0$. Thus, as one would anticipate based on the observation of the quantum discord in Fig. \ref{fig2}, even in the absence of entanglement, the average coherence that is completely stored on the correlations of the system is noticeably distinct from zero.

\section{Conclusions}

In summary, this paper explored the influence of magnetic anisotropies on the quantumness of a dipolar interacting magnetic system via a theoretical examination of the geometric quantum discord, measured by Schatten 1-norm, and the $l_1$ trace-norm quantum coherence. The analytical formulations for these quantum information quantifiers were obtained in terms of magnetic anisotropies. In this scenario, the effects of dipolar coupling constants on these quantifiers are highlighted. It is demonstrated that the presence of dipolar anisotropies increases the degree to which the system possesses quantum correlation and coherence.

As another remarkable result, it is proved that the global coherence, expressed in an arbitrary reference basis, determined by the co-latitude and longitude angles of the Bloch sphere representation, is totally stored within the correlations of the system. Moreover, according to the results, the behavior of quantum discord contains a notable hallmark of quantum level-crossing in the system, in contrast to the basis-dependent axial quantum coherence, which hides the energy-level crossover regarding the measured basis. 

Therefore, the dependency of the base on the quantum coherence specified by Baumgratz might be deleterious in identifying the crossing of levels owing to population changes originating from the changing of Boltzman weights due to the modification of the magnetic anisotropies of the studied system. In this regard, the average quantum coherence was measured numerically obtained in order to gain a viewpoint independent of the reference basis, unraveling that the average coherence is able to extract the signature of the energy-level crossover present in the measurement of quantum discord. 

Finally, the findings that were given provide light on the ways in which magnetic anisotropies caused by the dipolar interaction coupling of a dinuclear spin-1/2 system influence quantum correlations and coherence. Therefore, the dipolar interaction model is an excellent option for usage as a platform for quantum technologies that are based on quantum resources such as quantum coherence and quantum discord.

\section*{ACKNOWLEDGEMENTS}

C. Cruz gratefully acknowledges Mario Reis  for the valuable discussions. M. F. Anka thanks FAPERJ for financial support.


\end{document}